# LRSE: A Lightweight Efficient Searchable Encryption Scheme using Local and Global Representations

Ruihui Zhao
Waseda University
Japan
zachary@ruri.waseda.jp

Yuanliang Sun
Southeast University
China
sun254667307@gmail.com

Mizuho Iwaihara
Waseda University
Japan
iwaihara@waseda.jp

## ABSTRACT

Cloud computing is emerging as a revolutionary computing paradigm, while security and privacy become major concerns in the cloud scenario. For which Searchable Encryption (SE) technology is proposed to support efficient retrieval of encrypted data. However, the absence of lightweight ranked search with higher search quality in a harsh adversary model is still a typical shortage in existing SE schemes. In this paper, we propose a novel SE scheme called LRSE which firstly integrates machine learning methods into the framework of SE and combines local and global representations of encrypted cloud data to achieve the above design goals. In LRSE, we employ an improved secure kNN scheme to guarantee sufficient privacy protection. Our detailed security analysis shows that LRSE satisfies our formulated privacy requirements. Extensive experiments performed on benchmark datasets demonstrate that LRSE indeed achieves state-of-the-art search quality with lowest system cost.

## 1 INTRODUCTION

Cloud computing is a revolutionary computing paradigm which provides a flexible and economic strategy for data management and resource sharing [1], [2], thus is getting more and more attention from both academic and industry communities. However, security and privacy become major concerns in the cloud scenario when data owners outsource their private data onto public cloud servers to be accessed by the authenticated users. Usually, the cloud server is considered as curious and untrusted entities [3], thus there are risks of data exposure to a third party or even the cloud service provider itself. Therefore, providing sufficient security and privacy protections on sensitive data is extremely important, especially for those applications involved with health records, financial, government data, patents, managing passwords, private photos, etc. To avoid information leakage, the sensitive data should be encrypted before uploading onto the cloud servers, which makes it a big challenge to support efficient keyword based queries and rank the matching results on the encrypted data.

To address the issue, searchable encryption (SE) technology has been proposed in the literature in pursuit of search over encrypted data. For schemes [3,5] that realize flexible search, they only support Boolean keyword search or single keyword search and return inaccurate results that are often loosely related to the user's intent. In 2014, Cao et al. [6] firstly proposed an effective mechanism called MRSE to partially solve the multi-keyword ranked search problem according to the number of matching keywords between the query and documents, which established the foundation and basic framework of multi-keyword ranked search in the field of searchable encryption. As far as we know, most of latest schemes in SE follow this framework, such as [7, 8, 9, 11, 14, 15]. As a consequence, they have the same congenital drawbacks. Here we conclude the congenital drawbacks of existing searchable encryption schemes as follows:

1) Low search quality. Most of latest existing SE schemes which follow the classic MRSE framework [6] are based on keyword match method, which is functionally inferior in the view of current plaintext information retrieval and machine learning. For example, MRSE simply counts the number of matching keywords between query and documents and does not take the access frequencies of the keywords into account. Although following work such as [8, 9, 11, 14, 15] employs TF×IDF weighting to substitute occurrence bits in binary vectors, they are still too primal and basic schemes, because TF×IDF weighting is quite a rudimentary technique in information retrieval field.

2) Dimension disaster. In MRSE [6], the system overhead during the whole process of index construction, trapdoor generation, and executing the query, is mostly determined by matrix multiplication, in another word, the large dimension of the sparse vector according to the dictionary in this framework causes dimension disaster. This problem leads to a result that searchable encryption schemes cannot be practically put to use in a real-world scenario.

3) Lack of fuzzy search and intelligent search. For example, if data user inputs "producing" in the query, if the corresponding keyword in the dictionary is "produce", it is quite difficult for most existing SE schemes to return relevant results due to the lack of "producing" in the dictionary and keyword match based method. Let alone some more advanced examples in existing plaintext information retrieval, such as intelligent search using the semantic similarity, for example, "java" and "python", or "iPhone" and "cellphone". Because keyword match based method cannot measure semantic similarities in such cases.

On the other hand, in the branch of plaintext information retrieval (IR) and document filtering, such as a common practice in web search engines (e.g., Google search), data users tend to provide a query containing several keywords as the indicator of their search interest to retrieve the most relevant data. "Coordinate matching", i.e., as many matches as possible, and ranking matching documents by certain criteria has been widely used in the plaintext information retrieval (IR) field. In the field of document retrieval and natural language processing, machine learning based methods, such as word embeddings [4], are emerging as replacing traditional term vectors for measuring relatedness between terms. However, existing techniques in plaintext information retrieval and document filtering, such as [10] and [18], cannot be directly used in encrypted cloud scenario mentioned above for the reasons as follows: it is not practical for data owner to train word embeddings using CNN [24] or word2vec [4] for his local dataset each time, because the training process would greatly add to the time of building index. Besides, considering the local dataset of data owner may be small, it is not realistic to train a local model on a small dataset. In addition, in the framework of SE, the data owner and data user are separate entities, thus it's difficult for the data user to embed the query keywords into the same space of word embeddings as the data owner. In conclusion, how to choose and apply appropriate machine learning methods to the framework of SE remains a challenging task. To the best of our knowledge, in literature, there are no existing SE schemes based on machine learning methods and aiming to achieve the goal of higher search quality, lightweight search with low system cost and supporting fuzzy and intelligent search in a harsh adversary model. Our motivation and focus in this paper are achieving the above goals. In this paper, we propose a novel lightweight efficient multi-keyword ranked search over encrypted cloud data (LRSE) scheme that supports top-$k$ retrieval in a harsh adversary model. In summary, this paper makes the following contributions:

1) For the first time, it successfully introduces machine learning methods into SE framework. Our original idea is: we design the whole protocol to combine local and global representations of documents to guarantee the search quality.

2) It beats existing SE schemes on a benchmark dataset and achieves state-of-the-art search quality. Besides, it can also support fuzzy and intelligent search.

3) Experiments on a real-world dataset further show the proposed scheme introduces much lower overhead than existing SE schemes.

4) Thorough analysis investigating privacy issues in a harsh adversary model is given.

The remainder of this paper is organized as follows: we discuss existing related work on searchable encryption and plaintext document filtering based on machine learning methods in Section 2. In Section 3, we introduce the system model, adversary model, and security requirements. Section 4 describes the LRSE framework and proposed schemes, followed by Section 5, which focuses on security analysis. Section 6 presents search quality evaluation and system cost simulation results. At last, we conclude the paper in Section 7.

## 2 Related Work

### 2.1 Existing Searchable Encryption

Cao et al. [6] firstly propose an effective mechanism which can partially solve the multi-keyword ranked query problem according to the number of matching keywords in a vector space model (VSM), however, MRSE does not take the access frequencies of the keywords into account. It only returns the documents ordered by the number of matched keywords. Besides, MRSE has problems such as low search quality and dimension disaster. Yu et al. [11] propose a two-round searchable encryption that supports top-$k$ multi-keyword retrieval (TRSE) scheme, which can guarantee high security and practical search accuracy. Its main idea is combining VSM model and homomorphic encryption. The VSM model helps to provide search accuracy and homomorphic encryption guarantees high security. However, it is still based on keyword match method and has the problems of dimension disaster, low search quality and lack of intelligent search. Besides, homomorphic encryption usually leads to much larger system overhead. Zhao et al. [9] firstly propose a privacy-preserving personalized search (PPSE) scheme which supports personalized search in SE. The main idea is to combine TF×IDF weighting VSM model and preference weight formally generated in local user interest model by the data user. However, dimension disaster leads to huge system cost because of employing VSM model.

In 2016, Xia et al. [14] propose a secure and dynamic multi-keyword ranked search scheme over encrypted cloud data. The goal is supporting dynamic update operations on documents and sub-linear search time. The main idea is combining secure *kNN* encryption with TF×IDF weighting VSM model, constructing a special tree-based index structure and relying on a greedy depth-first search to achieve sub-linear search time. However, it's still based on TF×IDF keyword match method and suffers from low search quality and lack of fuzzy and intelligent search. Besides, time cost for index tree construction is huge, although its index building part costs sub-linear search time, the system cost is still not lightweight enough compared with our dimension reduction based scheme. In 2016, Li et al. [15] propose a scheme enabling fine-grained multi-keyword search supporting classified sub-dictionaries over encrypted cloud data. The goal is supporting complicated logic search operations of keywords, and reducing system cost. The main idea is combining secure *kNN* encryption and VSM model, introducing the TF×IDF relevance score and preference weight to improve search quality and realizing the "AND", "OR", "NO" operations in the multi-keyword search.

Besides, it employs the classified sub-dictionaries technique to reduce the system cost. However, it is still based on TF×IDF weighting keyword match based method. Besides, system cost is still much larger than our lightweight scheme realized by dimension reduction.

As to latest SE schemes in 2017, [16] focuses on proposing multi-level access control policy using broadcast encryption, which means different users should have different access rights, and only be permitted to search over data that they are authorized. [17] focuses on multi-user searchable encryption (MSE), which achieves fine-grained access control to grant and revoke the privileges of users without a trusted third party (TTP), and the key distribution is integrated with user authorization and search procedures. However, [16, 17] focus more on security protocols design and support novel functions such as multi-level access control and multi-user SE schemes, their search quality and system cost part have no improvement. While our main focus is improving search quality and achieving lightweight search.

## 2.2 Plaintext Document Filtering based on machine Learning

In 2016, Nalisnick et al. [10] propose a dual embedding space model (DESM), which uses dual word embeddings, one for generating query vector, and the other one for generating document vector. DESM can be used to calculate the similarity between a document and a query term, complementing the traditional term frequency based approach. In 2017, Mitra et al. [18] propose a duet model that jointly learns two deep neural networks using local and distributed representations of text, respectively. They train word embeddings on the local dataset and the training dataset using CNN, respectively, and add the two scores at the final step. Thus, it performs better than all the existing schemes that merely use traditional baselines or neural baselines, and achieves state-of-the-art search quality in the field of plaintext document ranking. However, [10] and [18] are not directly applicable in the context of encrypted cloud data retrieval. The reasons have been elaborated in the introduction part.

## 3 Problem Formulation

### 3.1 System Model

As illustrated in Fig. 1, our scheme involves three different entities.

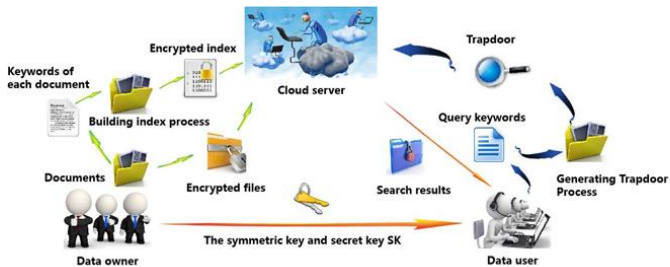

**Figure 1: System Model.**

1) **The cloud server:** the cloud server is an intermediate entity hosting third-party data storage and retrieval services to authenticated data users. When received a trapdoor from the data user, the cloud server will locate the matching documents by scanning the indexes $I$, calculate corresponding relevance scores, and return the ranked top-$k$ results to the data user.

2) **The data owner:** the data owner encrypts a collection of documents $D$ using symmetric key encryption algorithm and builds a searchable index $I$ according to the building index process, then he outsources both the encrypted indexes $I$ and encrypted files $C$ onto the cloud server. After that, the data owner sends $SK$ to the data user, and the details of $SK$ is stated in section *4.1*.

3) **The data user:** the data user generates a trapdoor with $SK$ and sends it to the cloud server. Afterward, the data user is returned the most relevant top-$k$ encrypted documents by the cloud server, then he decrypts and makes use of them with the help of $SK$.

Note here, there is a special case where the data owner is the same as the data user, that's to say, the data owner keeps his secret key $K$ by himself and only searches over his own encrypted documents. Then there are only two entities left in this special case: the data owner and the cloud server. This case also has many vivid applications in daily life, for example, when we upload our local documents in the phone and PC to the cloud server, such as Google, Amazon, Microsoft, and Baidu cloud drive, there is a horrible security problem: some documents related to passwords, patents to be published, private photos, health records, etc. are exposed to the mentioned companies. This special case could help solve the mentioned problem.

### 3.2 Adversary Model and Security Requirements

The cloud server is considered as honest but curious, i.e., it is designed to execute the service algorithm faithfully, however, it is also curious and eager to attain sensitive information. In this paper, we define all the security requirements in a harsh adversary model, which means *level 3* attack model in [13]. Let $H$ means the knowledge that cloud server knows. In level *3* attack model, cloud server observes a set of plain document vectors $P$ in $D$, and knows the corresponding encrypted index, i.e., $H = <E(D), P, I>$, where $P \subset D$, $I(t) = E_T(t, SK)$ $for\ all\ t \in P$. As described in [13], if an encryption scheme resists a higher-level attack, it resists a lower level one as well. Thus, in this paper we only prove our schemes could resist *level 3* attack. Security requirements are defined as follows:

1) Data, index, and trapdoor privacy: data privacy means that LRSE should prevent the cloud server from poking its nose into the outsourced data. Index privacy means that the index should be constructed to prevent the cloud server from performing association attack i.e., deducing any association between keywords and encrypted documents. Besides, trapdoor privacy means the keywords the user submits according to his interest is well protected by the complexity of trapdoor generating algorithm.

2) Unlinkability of trapdoors: as described in [6], in *level 3* adversary model, the cloud server is more powerful and possesses

some statistical information to carry out Scale Analysis Attack [6]. The linkability of trapdoor may cause leakage of privacy. For example, the cloud server may determine two trapdoors are originated from the same keywords or whether the trapdoor contains some certain keywords. We should assure that the cloud server would not be able to identify the keywords in a query even if some background information had been leaked.

## 4 The Design of LRSE

In this section, we firstly propose a basic idea for the LRSE by elegantly combining machine learning methods with an improved *kNN* scheme, which mainly consists of the following four phases: Initialization, BuildIndex, GenTrapdoor, and Query.

### 4.1 Initialization

**Data preprocessing.** The data owner executes data prepossessing process on his documents, including stop words, punctuations filtering, lowercase, tokenize, and lemmatize using the famous *NLTK* toolkits[1]. Afterward, the data owner extracts *top-10* keywords for each document $D_i$ in dataset *D* using *TF×IDF* weighting method as Equ.(1):

$$w(t_{ij}, D_i) = \frac{tf(t_{ij}, D_i) \cdot idf(t_{ij})}{\sqrt{\sum_{t_{ij} \in D_i}[tf(t_{ij}, D_i) \cdot idf(t_{ij})]^2}} \quad (1)$$

Where, $$idf(t_{ij}) = log\left(\frac{|D|}{|D_{t_{ij}}|} + 0.01\right) \quad (2)$$

In this equation, $tf(t_{ij}, D_i)$ is the term frequency of keyword $t_{ij}$ in document $D_i$. Then the data owner generates keyword-document matrix *A* using vector space model (*VSM*), where each item in matrix *A* is the the *TF×IDF* weight value $w(t_{ij}, D_i)$ of keyword $t_{ij}$.

Afterward, the data owner executes the reduce dimension representation by Singular Value Decomposition (*SVD*) [19, 20] as follows: for a keyword-document matrix *A* with *t* rows and *|D|* columns, *SVD* finds the approximation matrix called $A_{n_1}$ as:

$$A_{n_1} = U_{n_1} S_{n_1} V_{n_1}^T \quad (3)$$

Usually we take $n_1 \in [200, 500]$. In Equ.(3), *t* is the original keyword number in the dataset *D*, $n_1$ is the concept number after dimension reduction by *SVD*, $U_{n_1}$ is the $t \times n_1$ keyword-concept matrix, $S_{n_1}$ is the $n_1 \times n_1$ concept matrix, and $V_{n_1}^T$ is the $n_1 \times |D|$ concept-document matrix.

**Generate external word embeddings.** Word embeddings [4] are a generic name of a set of *NLP* techniques, where each unique word is represented by a relatively low dimension vector of real numbers. Their models are learned through two-layer neural networks to capture linguistic contexts of words. We trained word embeddings using word2vec [4] in this paper. Word2vec helps to represent each word *w* of the training set as a vector of features, where this vector is supposed to capture the contexts in which *w* appears.

In this paper, we choose the whole *2015* English Wikipedia corpus[2], containing *2,126,359* words, as the training set to train an external word embeddings for our LRSE framework. We denote the pre-trained word embeddings using the notation $W_e$. The training parameters of word2vec are set as follows: continuous bag of words (*CBOW*) model instead of the skip-gram (word2vec options: cbow = *1*); the output vectors size is usually set as $n_2 \in [50, 300]$, and we choose *100* for $n_2$ in this paper (word2vec options:size = *100*); the number of negative samples is set to *5* (word2vec options: negative = *5*).

**Generate key.** The data owner randomly generates an (*n+1*)-dimension binary vector as *S* and two (*n+1*)×(*n+1*) invertible matrices {$M_1$, $M_2$}, where $n = n_1 + n_2 \in [250, 800]$, and $n_1$ is the the concept number after dimension reduction by *SVD*, usually $n_1 \in [200, 500]$; $n_2$ is dimension of word vectors mapping from the pre-trained word embeddings, usually $n_2 \in [50, 300]$. The secret key *K* is in the form of a *3*-tuple as {*S*, $M1$, $M2$}. Then the data owner sends $SK = (K, sk, U_{n_1}, S_{n_1}, V, W_e)$ to data users through a secure channel. *sk* is the symmetric key (e.g. AES [14]) used to encrypt documents outsourced to cloud server; $U_{n_1}$ and $S_{n_1}$ are generated in *SVD* dimension reduction; *V* is the dataset vocabulary of the data owner; $W_e$ is the pre-trained word embeddings. We extend the dimension of vectors and matrices in our schemes to (*n+1*)-dimension because of the improved secure *kNN* scheme.

Note that *S* is part of secret key *K*, which is exactly a binary vector, acting as an indicating vector in the spilt process of building index and generating trapdoor. *S* is randomly generated by the data owner using existing Random Number Generation Algorithm (*PRGA*) in the field of information security, which is not within the scope of this paper.

### 4.2 BuildIndex

**Generate initial document vector.** For each document $D_i$, the data owner generates $d_1 = d^T U_{n_1} S_{n_1}^{-1}$, then calculates its standard unit vector: $\vec{d_1} = \frac{d_1}{\|d_1\|}$, where *d* is the $i-th$ column vector in matrix *A*, denoting the *TF×IDF* weighting document vector for document $D_i$. Then the data owner maps the extracted keywords for each document in word embeddings: let $d_2 = \sum_{t_{ij} \in D_i} d_{ij} \cdot w(t_{ij}, D_i)$, where $d_{ij}$ is the embedding vector for the *j-th* keyword of document $D_i$; $w(t_{ij}, D_i)$ is the normalized *TF×IDF* weight of keyword $t_{ij}$ in document $D_i$ in Equ.(1); then calculate its standard unit vector: $\vec{d_2} = \frac{d_2}{\|d_2\|}$. Afterward, the data owner generates the initial document vector for document $D_i$: $\vec{D_i} = (\vec{d_1}, \vec{d_2}, 1)$, which is an (*n + 1*) dimensional vector, and $n = n_1 + n_2$.

---

[1] http://www.nltk.org/

[2] https://dumps.wikimedia.org/enwiki/latest/enwiki-latest-pages-articles.xml.bz2

**Splitting process.** The data owner executes the splitting process using the secret key $K = \{S, M1, M2\}$, For $m=1$ to $n+1$, if $\vec{S}[m] = 1$, then $\overrightarrow{D_{i'}}[m]$ and $\overrightarrow{D_{i''}}[m]$ are set to two random numbers so that their sum is equal to $\overrightarrow{D_i}[m]$; else, $\overrightarrow{D_{i'}}[m]$ and $\overrightarrow{D_{i''}}[m]$ are set as the same as $\overrightarrow{D_i}[m]$. Every plaintext subindex $\overrightarrow{D_i}$ is then spilt into a document vector pair donated as $\{\overrightarrow{D_{i'}}, \overrightarrow{D_{i''}}\}$. Finally, the subindex $I_i = \{M_1^T \overrightarrow{D_{i'}}, M_2^T \overrightarrow{D_{i''}}\}$ is built for every encrypted document $C_i$. Let $I$ denote the set of subindex $I_i$, we call $I$ index in our scheme. Finally the data owner encrypts his documents using symmetric key $sk$, and outsources the encrypted documents $C$ together with index $I$ to the cloud server.

### 4.3 GenTrapdoor

**Generate initial query vector.** The data user inputs a set of query keywords according to his interest. Let $q$ denote the query vector according to the received vocabulary $V$. The data user generates $q_1 = q^T U_{n_1} S_{n_1}^{-1}$. Then calculate its standard unit vector: $\overrightarrow{q_1} = \frac{q_1}{\|q_1\|}$. Afterward, the data user maps the query keywords in word embeddings. Let $q_2 = \sum_{q_k \in Q} q_k$, where $q_k$ is the embedding vector for the *k-th* keyword in the query; then calculate its standard unit vector: $\overrightarrow{q_2} = \frac{q_2}{\|q_2\|}$. Afterward, the data user generates an initial vector $\vec{Q}$ for query $Q$ using $\vec{Q} = (\overrightarrow{q_1}, \overrightarrow{q_2})$. $\vec{Q}$ is extended to $(r\vec{Q}, t)$, which is *(n+1)-dimension*, $r$ and $t$ are random numbers, and $r > 0$. Note here only the data user knows the exact values of $r$ and $t$.

**Splitting process.** The splitting process in generating trapdoor is on the contrary to the splitting process in building index. For $m=1$ to $(n+1)$, if $\vec{S}[m] = 0$, $\overrightarrow{Q'}[m]$ and $\overrightarrow{Q''}[m]$ are set to two random numbers so that their sum is equal to $\vec{Q}[m]$; else, $\overrightarrow{Q'}[m]$ and $\overrightarrow{Q''}[m]$ are set the same as $\vec{Q}[m]$. Finally, the trapdoor $T$ is generated as $\{M_1^{-1}\overrightarrow{Q'}, M_2^{-1}\overrightarrow{Q''}\}$ for the query. Finally the data user sends the trapdoor $T$ to the cloud server, with which the cloud server could execute the query process.

### 4.4 Query

For each subindex $I_i$, with the trapdoor $T$, the cloud server computes the relevance scores as shown in the following equation, ranks all relevance scores and returns the top-*k* ranked encrypted documents to the data user.

$$\begin{aligned} Score(I_i, T) &= I_i \cdot T = \{M_1^T \overrightarrow{D_{i'}}, M_2^T \overrightarrow{D_{i''}}\} \cdot \{M_1^{-1}\overrightarrow{Q'}, M_2^{-1}\overrightarrow{Q''}\} \\ &= \overrightarrow{D_{i'}} \cdot \overrightarrow{Q'} + \overrightarrow{D_{i''}} \cdot \overrightarrow{Q''} \\ &= (\overrightarrow{D_i}, 1) \cdot (r\vec{Q}, t) \\ &= (\overrightarrow{d_1}, \overrightarrow{d_2}, 1) \cdot (r\overrightarrow{q_1}, r\overrightarrow{q_2}, t) \\ &= r\, Score_1(\overrightarrow{d_1}, \overrightarrow{q_1}) + r\, Score_2(\overrightarrow{d_2}, \overrightarrow{q_2}) + t \quad (4) \end{aligned}$$

where,

$$Score_1(\overrightarrow{d_1}, \overrightarrow{q_1}) = \frac{d^T U_{n_1} S_{n_1}^{-1} \cdot q^T U_{n_1} S_{n_1}^{-1}}{\|d^T U_{n_1} S_{n_1}^{-1}\| \cdot \|q^T U_{n_1} S_{n_1}^{-1}\|}$$

$$Score_2(\overrightarrow{d_2}, \overrightarrow{q_2}) = \frac{\sum_{t_{ij} \in D_i} d_{ij} \cdot w(t_{ij}, D_i) \cdot \sum_{q_k \in Q} q_k}{\|\sum_{t_{ij} \in D_i} d_{ij} \cdot w(t_{ij}, D_i)\| \cdot \|\sum_{q_k \in Q} q_k\|}$$

## 5 Security Analysis

In this section, we analyze the security properties in a *level 3* adversary model as illustrated in section *3.2*. We will focus on three aspects: data privacy, index and trapdoor privacy, and trapdoor unlinkability.

### 5.1 Data Privacy

Traditional symmetric key encryption techniques (e.g., *AES* [12]) could be properly utilized here to guarantee data privacy and is not within the scope of this paper.

### 5.2 Index and Trapdoor Privacy

We refer to the original secure kNN scheme [13] and give THEOREM 1 as follows:

*THEOREM 1. LRSE is resilient to a level-3 attack if the bit string S in secret key K is kept confidential.*

*Analysis.* The proof process is based on the same principle as *THEOREM 6* in the original secure *kNN* scheme [13]. Thus we do not prove it again in our paper. The only differences exist in our added artificial attributes: in the process of building index of [13], the $(n + 1)$-dimension is set to $-0.5\|p\|^2$, while the $(n + 1)$-dimension is set to *1* as described in section 4.3 in our scheme; in the process of generating trapdoor, the first $n$ dimensions are given by $r\vec{Q}$, and the $(n + 1)$-dimension is also set to $r$, while in our scheme we set the $(n + 1)$-dimension as a random number $t$. Thus our vector dimension and encryption procedure including the splitting process are the same as the original secure *kNN* scheme. Actually the core idea in [13] is: if the cloud server does not know the splitting configuration, i.e. the bit string *S*, the equations he models for solving the key $K$ is $M_1^T \hat{p}' = p'$, where the cloud server only knows the details of $p'$, while he does not know $M_1^T$ and $\hat{p}'$, therefore he cannot decrypt the transformation matrix $M_1^T$ in the key $K$ due to lack of sufficient information. Thus the proof has nothing to do with the added artificial attributes, and only determined by the splitting process. Therefore the original proof process in [13] can be applied to our case. In another word, if the cloud server cannot derive the splitting configuration, i.e. the bit string *S* in the key *K* in LRSE scheme, the index and trapdoor privacy can be achieved.

Another issue is what the value of $n'$ would be sufficiently large. As stated in [13], the general consensus is that 1024-bit *RSA* [22] keys are roughly equivalent in strength to that of *80-bit* symmetric keys. As $n' \in (251, 801)$ in our scheme, which is much larger than *80*, so that the search space is sufficiently large. When $n'$ increases, the search space enlarges exponentially.

Besides, in the initial document vector and query vector, $\overrightarrow{d_2}$ and $\overrightarrow{q_2}$ are generated by computations on word embeddings, which may pose a new security challenge. Because attackers may know word embeddings trained on popular corpus such as Wikipedia and news streams. In fact, this attack example conforms to the *level 2* adversary model in [13]: cloud server knows a set of plain document vectors $P$; but does not know the corresponding encrypted index in $E(D)$, i.e., $H = <E(D), P>$ where $P \subset D$. As stated in [13], if an encryption scheme resists a higher level attack, it resists a lower level one as well. Thus our LRSE scheme can resist this kind of attack.

### 5.3 Trapdoor Unlinkability

The trapdoor should be constructed to prevent the cloud server from deducing the relationships of any given trapdoors and the corresponding keywords.

Table 1. The structure of $\vec{Q}$

| $\vec{Q}[1] \cdots \vec{Q}[n_1]$ | $\vec{Q}[n_1+1] \cdots \vec{Q}[n_1+n_2]$ | $\vec{Q}[n_1+n_2+1]$ |
|---|---|---|
| $n_1$-dimension | $n_2$-dimension | 1-dimension |
| $r \cdot \vec{q_1}$ | $r \cdot \vec{q_2}$ | $t$ |

According to the procedure of generating trapdoor in section *4.3*, the details of $\vec{Q}$ is illustrated as Tab.*1*, where $r$ and $t$ are random numbers defined by data user, and $r > 0$. $\vec{q_1} = \frac{q^T U_{n_1} S_{n_1}^{-1}}{\|q^T U_{n_1} S_{n_1}^{-1}\|}$, and $\vec{q_2} = \frac{\sum_{q_k \in Q} q_k}{\|\sum_{q_k \in Q} q_k\|}$. Finally, $\vec{Q}$ is split to $(\vec{Q'}, \vec{Q''})$ according to the splitting indicator $S$ in the splitting process. Specifically, for $m = 1$ to $(n+1)$: if $\vec{S}[m] = 0$, then $\vec{Q'}[m]$ and $\vec{Q''}[m]$ are set to two random numbers so that their sum is equal to $\vec{Q}[m]$; assume in $S$ the number of 0 is $\mu$, and each dimension in $\vec{Q}$ is $\delta_q$-bits. Random numbers $r$ and $t$ are $\delta_r$-bits and $\delta_t$-bits respectively. Besides, assume there are $\alpha$ sequences composed of the values of all dimensions in $\vec{Q}$. Note that $\mu$, $\delta_q$, $\delta_r$ and $\delta_t$ are independent of each other. Then in our LRSE scheme, we can compute the probability that two trapdoors are generated the same as follows:

$$P = \frac{1}{\alpha \cdot 2^{\delta_r} \cdot (2^{\delta_q})^\mu \cdot 2^{\delta_t}} = \frac{1}{\alpha \cdot 2^{\delta_r + \delta_t + \mu \cdot \delta_q}} \quad (5)$$

Therefore, the larger values of $\mu$, $\delta_q$, $\delta_r$ and $\delta_t$, the lower probability that two trapdoors are the same, for example, if we choose 1024-bit Random number $r$, $\delta_r = 1024$, then $P < 1/2^{1024}$. As a result, the probability that two trapdoors are generated the same can be negligible.

The above argument is just showing the size of possible space when generating trapdoors. However, the cloud may execute some attack based on his obtained background information and powerful computation ability. For example, there is a kind of *Scale Analysis Attack* introduced in [6]. Basically, in *MRSE* schemes, vectors in the original document and query vectors are sparse binary vectors. Thus in its query equation the values of $\vec{D} \cdot \vec{Q}$ equals to the number of matched keywords. Then the cloud server can deduce whether a trapdoor contains some certain keywords by constructing some equations as shown in [6]. However, our scheme can resist *Scale Analysis Attack* because the initial vectors in our schemes are low-dimensional vectors of real numbers and the values of $\vec{D_i} \cdot \vec{Q} = Score_1(\vec{d_1}, \vec{q_1}) + Score_2(\vec{d_2}, \vec{q_2})$ in the Equ. (4), which represent the relevance scores of document $D_i$ and query $Q$, disclose no relationship between the trapdoor and certain keywords.

**Discussion 1.** Here we must admit that our scheme also cannot protect against Access Pattern [6] as most existing SE schemes such as [3][6][8][9][11][14][15][23], which is defined as the sequence of ranked search results. For example, the cloud server could conduct frequency analysis on the access pattern and deduce the linkability information of two trapdoors. Our proposed scheme is not designed to protect against Access Pattern, because of the efficiency consideration just as most existing SE schemes (excluding costly PIR technique [3]).

## 6 Performance Evaluation

### 6.1 Search quality experiments

**Experiment settings.** Most existing SE schemes such as [6][9][15][23] conduct their search quality experiments on a real-world dataset: the *NSF research award dataset*[3] or *Enron Email Dataset*[4] using the criteria of *P@K*. However, these datasets only give the keyword *ID* of each document, while there is no given queries and gold standard results for each query. Besides, the criteria of *P@K* ignores the ranking order of retrieval results and their relevance scores to the query. In our paper, we employ the *Cranfield dataset*[5] as the benchmark dataset, which is a pioneering test collection in the field of information retrieval. The details of *Cranfield* are summarized as follows: the titles and articles are collected from aerodynamics journal articles in the *UK*; there are *1,400* documents and *225* queries; in the file named "cranqrel", for each query they have given the relevance assessments of each document, which range from *0* (irrelevant) to *4* (perfectly relevant). We have implemented some most recent SE schemes as our baselines. In this paper, we use the criteria of *NDCG@3, 10*, which considers the ranking orders and relevance scores of retrieval results, and is defined as follows:

$$NDCG_k = DCG_k / IDCG_k \quad (6)$$

Where, $DCG_k = \sum_{i=1}^{k} \frac{rel_i}{log_2(i+1)}$, $IDCG_k = \sum_{i=1}^{|REL|} \frac{rel_i}{log_2(i+1)}$

*|REL|* represents the list of relevant documents (ordered by their relevance scores) in the returned results up to position *k* (i.e. *3* and *10*, respectively).

**Search quality results.** Tab.*2* reports *NDCG* based evaluation results on the benchmark dataset for our LRSE scheme and all the baseline schemes.

---

[3] https://kdd.ics.uci.edu/databases/nsfabs/nsfawards.html
[4] https://www.cs.cmu.edu/~./enron/
[5] http://ir.dcs.gla.ac.uk/resources/test_collections/cran/

**Table 2. NDCG results of existing SE schemes and our LRSE scheme**

| SE schemes | NDCG@3 | NDCG@10 |
|---|---|---|
| MRSE_I (keyword-count based) [6] | 0.434 | 0.483 |
| MRSE_II (keyword-count based) [6] | 0.351 | 0.400 |
| TF×IDF-based [8][9][11][14][15] | 0.655 | 0.666 |
| DESM-based(IN-IN) [10] | 0.147 | 0.199 |
| DESM-based(IN-OUT) [10] | 0.218 | 0.285 |
| SVD only (Our scheme) | 0.633 | 0.651 |
| TF×IDF weighting Wiki embeddings only (Our Scheme) | 0.474 | 0.521 |
| **LRSE (Our scheme)** | **0.672** | **0.673** |

From Tab.2 we can observe that LRSE achieves state-of-the-art search quality and beats existing SE schemes on *NDCG* scores including *TF×IDF*-based schemes such as [8][9][11][14][15]. In the implement of LRSE, we choose the concept number after dimension reduction by *SVD* as $n_1 = 300$, and the vector dimension in word embeddings as $n_2 = 100$. The MRSE schemes are based on simply counting the matched keywords number among the index and trapdoor, which is quite an original keyword match based method. Let alone MRSE_II scheme adds *U* dummy keywords to the final results, thus we can see MRSE_II lose to MRSE_I because of the inserted dummy keywords. Here we executed MRSE_II scheme using the following parameters as written in [6]: $U = 200, V = 100, \sigma = 0.5, \mu = 0, c = 0.87$, thus each dummy keywords $\varepsilon^{(j)}$ follows the same uniform distribution $M(-0.87, 0.87)$, where $j \in [1,100]$. We also integrate DESM model [10] which uses external dual word embeddings[6] trained by Microsoft on Bing queries into SE framework as one of our baselines. In our scheme, vector $d_2$ employs *TF×IDF* weighting word vectors mapping from Wikipedia embeddings, which is different from the formulas representing documents and query vector in DESM paper. From Tab.2 we can see that DESM-based schemes performs much worse than our scheme using only *TF×IDF* weighting Wikipedia embeddings, because DESM relies on external word embeddings trained on Bing queries, not on the test dataset, while the *cranfield* dataset is relatively old, and only focuses on glossary in aerodynamics journal articles, which is quite different from Bing queries.

In literature of the plaintext document filtering, *SVD* only is expected to perform better than *TF×IDF*, because the dimension reduction results as semantic space, and helps ignore some noise. However, in the cranfield dataset, explicit keyword match of aerodynamics glossary is dominant. Thus our *SVD*-based scheme slightly loses to *TF×IDF* because dimension reduction of *SVD* causes information loss, and its function of latent semantic relationship contribute little in this dataset. Therefore it calls for the help of external word embeddings in LRSE scheme. We can see that our complete LRSE scheme, the combination of SVD and TF×IDF weighting Wikipedia embeddings beats *TF×IDF*-based schemes [8, 9, 11, 14, 15], because *TF×IDF* weighting external word embeddings adds global semantic information and *TF×IDF* factor of local keywords in the final search results, which contributes to the search results.

---
[6] https://www.microsoft.com/en-us/download/details.aspx?id=52597

In addition, our scheme can also partly support fuzzy search and intelligent search due to the help of *NLTK* toolkits[1] in the data preprocessing process. Besides, *SVD* concept space and word embeddings can help measure semantic relatedness between keywords, for example, if we search "Java.", the top-*k* keywords are as follows: "Java", "swing", "android", "c++", "c#", "python", "eclipse", etc. And if we input "android" as a query keyword, the top-*k* relevant keywords are "Android", "iPhone", "blackberry", "emulator", "java", etc.

**Discussion 2.** In this paper, what we provide is a general framework. For example, when choosing external word embeddings, the data owner can replace our provided Wikipedia word embeddings using the dual word embeddings[6] if the documents in his dataset are relatively new and more relevant to Bing queries. If so, attackers can also obtain the external word embeddings, and the security issues have been discussed in section *5.2*.

**Discussion 3.** Our scheme is expected to perform even better than *TF×IDF*-based SE schemes on more common datasets that are not so old and made up of the dominant glossary. Because in that case, our *SVD* and external word embeddings would contribute more by measuring the latent semantic distance of keywords and introducing global information. In the future, we need do experiments on more datasets to prove the generality.

**Discussion 4.** LRSE also lightens the problems caused by dictionary updates in most existing SE schemes: since the vocabulary size of our word embeddings is large enough, we almost do not need to change the vocabulary. Besides, LRSE can also reduce the system cost even when the vocabulary in our scheme has to be updated, through dimension reduction by utilizing *SVD* and word embeddings.

**Discussion 5.** Actually our scheme can also support personalized search. In the process of generating trapdoor, we can refer to Refs. [8][15] to make the data user assign a number as his preference weight, or refer to Ref. [9] to build a local user interest model to formally generate user preference weight. If so, when generate trapdoor in section 4.3, the keyword vector should firstly multiply its preference weight. Because we did not find a benchmark dataset containing gold standard for personalized search, we did not discuss the details of personalized search in the evaluation part in this paper.

### 6.2 System cost experiments

We conduct a thorough experiment on a real-world dataset: the *NSF research award dataset[3]* and evaluate the performance of LRSE compared with existing schemes [6, 9, 15, 23]. The details of benchmark dataset are summarized as follows: (a) *129,000* abstracts describing NSF awards for basic research, (b) bag-of-word data documents extracted from the abstracts, (c) a list of

keywords for indexing the bag-of-word data in the file named "1wk1wd", from which we randomly select numbers of documents and conduct real-world system cost experiments on a computer using Intel(R) Core(TM) *i5-4590* CPU *3.30* GHz with *16.0* GB RAM. The criteria is time cost (*s*) or space cost (*KB*). Note in the implement of LRSE scheme, we choose vector dimension *n* as *300* to *700* for the dictionary size of *2,000* to *10,000*, respectively.

**1) *Building Index:*** as depicted in *Fig.2(a)* and *Fig.2(b)*, the time cost of building index is determined by the number of documents and the computation complex of building each subindex, and LRSE achieves much more lightweight system cost than existing schemes. As depicted in section *4.2*, the major computation of building a subindex includes four parts: search in word embeddings for each keyword of the document, generate the initial vector using mentioned equations, the splitting process and two multiplications of an ($n + 1$) × ($n + 1$) matrix and an ($n + 1$)-dimension vector. As illustrated in *Fig.2(a)*, given the same size of *1,000* documents with different vector dimensions, the index construction time of *LRSE* is much less than baseline schemes due to the difference of their vector dimensions. In the implement of FMSCS [15], we divide the total dictionary into *1* common sub-dictionary and *3* professional sub-dictionaries to generate index as illustrated in [15]. Although time cost of FMSCS is much less than other existing schemes, our LRSE scheme is even more lightweight because of dimension reduction. *Fig.2(b)* shows the time cost of building index with different document number when the vector dimension in LRSE is *300*, while the keyword number in the dictionary is *4,000* in all the schemes, we can see the time cost of LRSE scheme is much less. The time cost of building the whole index is almost linear with the size of dataset since the time cost of building each subindex is fixed. *Fig.2(b)* shows time cost of building index with different document number, the keywords number in the dictionary is *4,000*, and we specify vector dimension in LRSE as *400*. We can observe that the time cost is almost linear with the document number. Besides, in *Fig.3* we compare the storage overhead of subindex in *LRSE, MRSE_I, PPSE, etc.* within different sizes of vector dimension. The size of subindex is absolutely linear with the size of vector dimension.

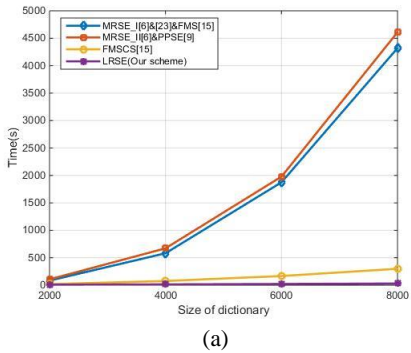

(a)

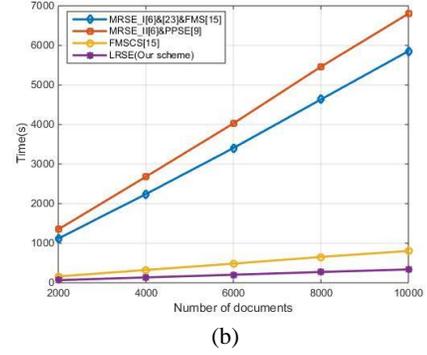

(b)

**Figure 2: (a)** Time cost of building index with different vector dimension using 1,000 documents; **(b)** Time cost of building index with different document number.

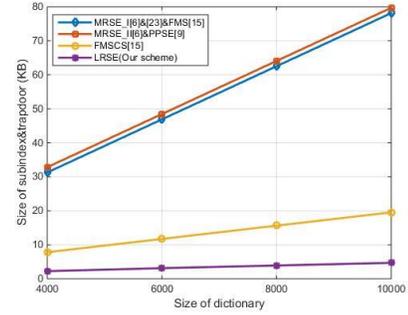

**Figure 3:** Size of subindex/trapdoor.

**2) *Generating Trapdoor:*** time cost of generating each trapdoor is determined by four parts: search over word embeddings for each keyword in the query using *HashMap*, generate the initial vector using mentioned equations in section *4.3*, the complexity of the splitting process and multiplications of a matrix and two spilt query vector. As shown in *Fig.4(a)*, the time of generating a trapdoor is greatly affected by the dimension of vectors, and the trapdoor generating time of LRSE is much less than other schemes due to the difference of their vector dimensions. Besides, as illustrated in *Fig.4(b)*, the number of keywords in the query has little influence upon the results because the dimension of vector and matrices is always fixed, here the vector dimension in LRSE is set to *400* by the dimension reduction of LRSE, while the keyword number in the dictionary of all the schemes is set to *4,000*. With respect to the size of the trapdoor, it occupies the same space overhead as that of each subindex listed in *Fig.3*, which is only determined by the vector dimensions.

**3) *Executing Query:*** the major computation to execute a query in the cloud server consists of computing the similarity scores of each subindex and trapdoor, ranking similarity scores for all documents in the dataset and selecting top-*k* results from all the scored documents. From *Fig.5*, we can observe that the query time of LRSE is much less than other schemes due to the difference of vector dimensions. We choose *1,000* documents and set *k* to *50* in our experiment. We can learn that the query time is almost linear with the vector dimension.

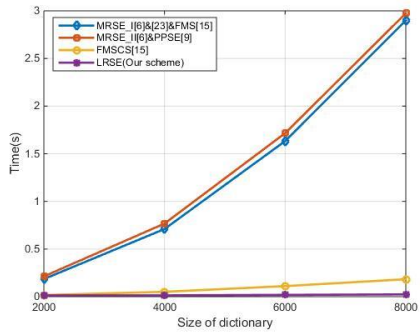

(a)

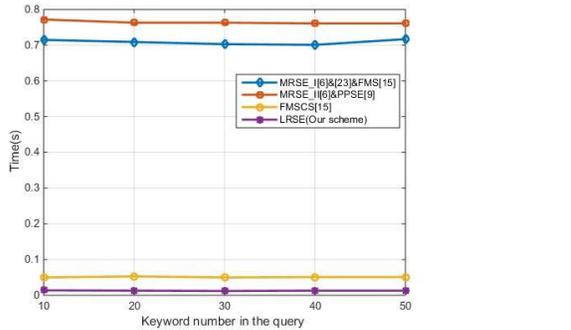

(b)

**Figure 4: (a) Time cost of generating trapdoor with different vector dimension; (b) Time cost of generating trapdoor with different keyword number in query.**

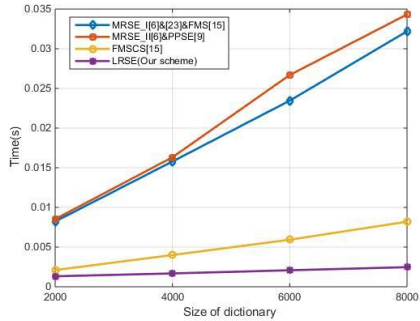

**Figure 5: Time cost of query with 1000 documents.**

## 7   Conclusion

In this paper, we propose a novel Lightweight Efficient Multi-keyword Ranked Search over Encrypted Cloud Data combining *SVD* and *TF×IDF* weighting word embeddings, which supports top-*k* retrieval in a harsh adversary model. For the first time, we introduce machine learning based method into SE framework, which provides search results with higher search quality. Besides, system cost experiments on a real-world dataset show LRSE is much more lightweight than existing SE schemes. We show examples that our scheme can partly support fuzzy and intelligent search. Security analysis shows that the proposed scheme accords with our formulated privacy requirements. In the future, we will conduct experiments on several more common datasets to verify the generality of search quality results.

## REFERENCES

[1] H. Li, Y. Dai, L. Tian, and H. Yang, "Identity-based authentication for cloud computing," in Proceedings of Cloud computing. Springer, 2009, pp. 157–166.
[2] H. Liang, L. X. Cai, D. Huang, X. Shen, and D. Peng, "A smdpbased service model for interdomain resource allocation in mobile cloud networks," IEEE Transactions on Vehicular Technology, vol. 61, no. 5, pp. 2222–2232, 2012.
[3] C. Wang, N. Cao, J. Li, K. Ren, and W. Lou, "Secure ranked keyword search over encrypted cloud data," in the 30th International Conference on Distributed Computing Systems (ICDCS). IEEE, 2010, pp. 253–262.
[4] Mikolov, Tomas, et al. "Distributed representations of words and phrases and their compositionality." Advances in neural information processing systems. 2013.
[5] J. Li, Q. Wang, C. Wang, N. Cao, K. Ren, and W. Lou, "Fuzzy keyword search over encrypted data in cloud computing," in Proceedings of INFOCOM. IEEE, 2010, pp. 1–5.
[6] N. Cao, C. Wang, M. Li, K. Ren, and W. Lou, "Privacy-preserving multi-keyword ranked search over encrypted cloud data," IEEE Transactions on Parallel and Distributed Systems, vol. 25, no. 1, pp. 222–233, 2014.
[7] Ishai Y, Kushilevitz E, Ostrovsky R, et al. Cryptography from Anonymity.[J]. Foundations of Computer Science Annual Symposium on, 2006, 2006:239-248.
[8] Shen Z, Shu J, Xue W. Preferred keyword search over encrypted data in cloud computing[C]. International Symposium on Quality of Service. ACM, 2013:1-6.
[9] Zhao R, Li H, Yang Y, et al. Privacy-preserving personalized search over encrypted cloud data supporting multi-keyword ranking[C]. Sixth International Conference on Wireless Communications and Signal Processing. IEEE, 2014:1-6.
[10] Nalisnick, Eric, et al. "Improving document ranking with dual word embeddings." Proceedings of the 25th International Conference Companion on World Wide Web. International World Wide Web Conferences Steering Committee, 2016.
[11] J. Yu, P. Lu, Y. Zhu, G. Xue, and M. Li, "Towards secure multi-keyword top-k retrieval over encrypted cloud data," IEEE Transactions on Dependable and Secure Computing, vol. 10, no. 4, pp. 239–250, 2013.
[12] N. Ferguson, R. Schroeppel, and D. Whiting, "A simple algebraic representation of Rijndael," in Selected Areas in Cryptography. Springer, 2001, pp. 103–111.
[13] W. K. Wong, D. W.-l. Cheung, B. Kao, and N. Mamoulis, "Secure knn computation on encrypted databases," in the 2009 ACM SIGMOD International Conference on Management of data. ACM, 2009, pp. 139–152.
[14] Xia, Zhihua, et al. "A secure and dynamic multi-keyword ranked search scheme over encrypted cloud data." IEEE Transactions on Parallel and Distributed Systems 27.2 (2016): 340-352.
[15] Li, Hongwei, et al. "Enabling fine-grained multi-keyword search supporting classified sub-dictionaries over encrypted cloud data." IEEE Transactions on Dependable and Secure Computing 13.3 (2016): 312-325.
[16] Alderman, James, Keith M. Martin, and Sarah Louise Renwick. Multi-level Access in Searchable Symmetric Encryption. IACR Cryptology ePrint Archive, Report 2017/211, 2017.
[17] Li, Zhen, et al. "Multi-user searchable encryption with a designated server." Annals of Tele-communications (2017): 1-13.
[18] Mitra, Bhaskar, Fernando Diaz, and Nick Craswell. "Learning to Match Using Local and Distributed Representations of Text for Web Search." Proceedings of the 25th International Conference Companion on World Wide Web. International World Wide Web (WWW) Conferences Steering Committee, 2017.
[19] Deerwester, S., Dumais, S. T., Furnas, G. W., Landauer, T. K., & Harshman, R. (1990). Indexing by latent semantic analysis. Journal of the American society for information science, 41(6), 391.
[20] Evangelopoulos, Nicholas E. "Latent semantic analysis." Wiley Interdisciplinary Reviews:   Cognitive Science 4.6 (2013): 683-692. a.
[21] O'Grady, John G., et al. "Controlled trials of charcoal hemoperfusion and prognostic factors in fulminant hepatic failure." Gastroenterology 94.5 (1988): 1186-1192.
[22] Rivest, Ronald L., Adi Shamir, and Leonard Adleman. "A method for obtaining digital sig-natures and public-key cryptosystems." Communications of the ACM21.2 (1978): 120-126.
[23] Sun, Wenhai, et al. "Verifiable privacy-preserving multi-keyword text search in the cloud supporting similarity-based ranking." IEEE Transactions on Parallel and Distributed Systems 25.11 (2014): 3025-3035.
[24] Krizhevsky, Alex, Ilya Sutskever, and Geoffrey E. Hinton. "Imagenet classification with deep convolutional neural networks." Advances in neural information processing systems. 2012.